%% file: main.tex
\documentclass[acmsmall,screen,nonacm]{acmart}
\AtBeginDocument{%
  }

\setcopyright{acmlicensed}
\copyrightyear{2026}
\acmYear{2026}
\acmDOI{XXXXXXX.XXXXXXX}
\acmConference[Conference acronym 'XX]{Make sure to enter the correct
  conference title from your rights confirmation email}{June 03--05,
  2018}{Woodstock, NY}
\acmISBN{978-1-4503-XXXX-X/2018/06}

\usepackage{algorithmic}
\usepackage{graphicx}
\usepackage{textcomp}
\usepackage{xcolor}
\usepackage{xspace}
\usepackage{booktabs}
\usepackage{multirow}
\usepackage{multicol}
\usepackage{varwidth}
\usepackage{hhline}
\usepackage{hyperref}
\usepackage{subcaption}
\usepackage{caption}
\usepackage{color}
\usepackage[T1]{fontenc}
\usepackage{tikz}
\usepackage{tikz-dependency}
\usepackage{enumitem}
\usepackage{tcolorbox}
\usepackage{amsmath}
\usepackage{amssymb}
\usepackage{adjustbox}
\usepackage{natbib}
\usepackage{algorithm}
\usepackage{booktabs}
\usepackage{pifont}
\usepackage{makecell}
\usepackage{xspace}
\usepackage{graphicx}
\usepackage{caption}
\usepackage{subcaption}

\newcommand{\ourmethod}{\textit{XSearch}\xspace}
\definecolor{zicong}{named}{green}

\definecolor{SoftRed}{RGB}{255, 228, 225}
\definecolor{SoftGreen}{RGB}{230, 255, 230}
\definecolor{SoftOrange}{RGB}{255, 229, 204}
\definecolor{SoftPurple}{RGB}{224, 224, 248}
\definecolor{SoftBlue}{RGB}{202, 240, 255}
\definecolor{SoftGray}{RGB}{237, 237, 237}
\definecolor{SoftYellow}{RGB}{255, 255, 204}
\colorlet{punct}{red!60!black}
\definecolor{delim}{RGB}{20,105,176}
\colorlet{numb}{magenta!60!black}
\definecolor{background}{RGB}{245,245,245}    
\definecolor{commentcolor}{RGB}{0,128,0}      
\definecolor{stringcolor}{RGB}{255,140,0}     
\definecolor{keywordcolor}{RGB}{0,0,255}      

\begin{document}

\title{XSearch: Explainable Code Search via Concept-to-Code Alignment}

\author{Yiming Liu}
\authornote{Both authors contributed equally to this research.}
\affiliation{%
  \institution{Shanghai Jiao Tong University}
  \city{Shanghai}
  \country{China}
}
\affiliation{%
  \institution{Shanghai Innovation Institute}
  \city{Shanghai}
  \country{China}
}
\email{liu_yiming@sjtu.edu.cn}

\author{Ruofan Liu}
\authornotemark[1]
\affiliation{%
  \institution{National University of Singapore}
  \country{Singapore}
}
\email{liu.ruofan16@u.nus.edu}

\author{Yun Lin}
\authornote{Corresponding author.}
\affiliation{%
  \institution{Shanghai Jiao Tong University}
  \city{Shanghai}
  \country{China}
}
\email{lin_yun@sjtu.edu.cn}

\author{Zicong Zhang}
\affiliation{%
  \institution{Shanghai Jiao Tong University}
  \city{Shanghai}
  \country{China}
}
\email{zhangzico@sjtu.edu.cn}

\author{Weiyu Kong}
\affiliation{%
  \institution{Shanghai Jiao Tong University}
  \city{Shanghai}
  \country{China}
}
\email{kwy160034@sjtu.edu.cn}

\author{Pengnian Qi}
\affiliation{%
  \institution{Huawei Technologies Co., Ltd}
  \city{Shenzhen}
  \country{China}
}
\email{qipengnian@huawei.com}

\author{Xiao Cheng}
\affiliation{%
  \institution{Huawei Technologies Co., Ltd}
  \city{Shenzhen}
  \country{China}
}
\email{chengxiao5@huawei.com}

\author{Weinan Zhang}
\affiliation{%
  \institution{Shanghai Jiao Tong University}
  \city{Shanghai}
  \country{China}
}
\affiliation{%
  \institution{Shanghai Innovation Institute}
  \city{Shanghai}
  \country{China}
}
\email{wnzhang@sjtu.edu.cn}

\author{Qianxiang Wang}
\affiliation{%
  \institution{Huawei Technologies Co., Ltd}
  \city{Shenzhen}
  \country{China}
}
\email{wangqianxiang@huawei.com}

\author{Linpeng Huang}
\affiliation{%
  \institution{Shanghai Jiao Tong University}
  \city{Shanghai}
  \country{China}
}
\email{lphuang@sjtu.edu.cn}

\renewcommand{\shortauthors}{Liu et al.}

\begin{abstract}
With the emergence of deep learning, semantic code search has been widely adopted in both academia and industry. 
These approaches embed natural-language queries and code snippets into a shared embedding space and retrieve results based on vector similarity. 
Despite their strong performance on benchmark datasets, they often suffer from poor explainability and generalization. 
Retrieved code may appear semantically similar yet miss critical functional requirements of the query, while providing no explanation of why the result was retrieved. 
Moreover, such failures become more severe under distribution shift, where models struggle to generalize to unseen benchmarks.

In this work, we propose \ourmethod, an intrinsically explainable code search framework.
Our key insight is that, by relying on global embedding similarity, all existing retrievers inherently take an inductive view.
They learn statistical patterns, rather than truly understand the query's functional requirements.
Therefore, we address the problem by reformulating code search as a deductive concept alignment problem.
At a high level, \ourmethod 
(i) identifies functional concepts in the query and 
(ii) explicitly aligns them with corresponding code statements.
This explain-then-predict design not only produces inherent concept-level explanations, 
but also mitigates shortcut learning that harms out-of-distribution generalization.
We train an encoder with explicit concept-alignment objectives and perform retrieval through explicit matching between query concepts and code statements. 
Experiments show that, when trained on CodeSearchNet with a small model size (GraphCodeBERT with 125M parameters), 
\ourmethod improves performance on out-of-distribution benchmarks from 0.02 to 0.33 (15×) over eight state-of-the-art retrievers, and consistently outperforms both encoder-based and decoder-based baselines with up to 7B parameters. 
A controlled user study further demonstrates that concept-alignment explanations enable users to accept or reject retrieved results both faster and more accurately.
\end{abstract}

\ccsdesc[500]{Software and its engineering~Search-based software engineering}
\ccsdesc[500]{Computing methodologies~Information extraction}

\keywords{Explainable Code Search, AI for Analysis and Testing}
  
\maketitle

\input{intro}

\input{motivating_example}

\input{approach}

\input{experiments/main}

\input{threat}

\input{conclusion}
\newpage
\section*{Data Availability}
The datasets and code used in this study are available at \url{https://anonymous.4open.science/r/Xsearch-2EC0}. 

\bibliographystyle{ACM-Reference-Format}
\bibliography{main}

\end{document}

%% file: intro.tex
\section{Introduction}

Code search is a fundamental activity in modern software development,
consuming up to 50\% of the developers' time during programming tasks such as code reuse, debugging, and feature implementation \cite{piorkowski2016foraging, xia2017measuring}.
Modern code search solutions have been shifted from keyword-based lexical matching techniques \cite{chan2012searching,mcmillan2011portfolio,lv2015codehow, nie2016query,hill2014nl,lu2015query,raghothaman2016swim} to semantic retrieval models \cite{gu2018deep, husain2019codesearchnet, feng2020codebert, guo2020graphcodebert, guo2022unixcoder, kim2018facoy, saieva2024reinforest}.
By mapping queries and code into a shared dense vector space, these models capture high-level semantic similarities.
However, despite their success on in-distribution benchmark datasets, 
state-of-the-art retrievers remain black boxes, and often fail in real-world, out-of-distribution scenarios.
We identify two fundamental challenges in existing code search frameworks.

\begin{figure*}[t]
  \centering
  \includegraphics[width=\linewidth]{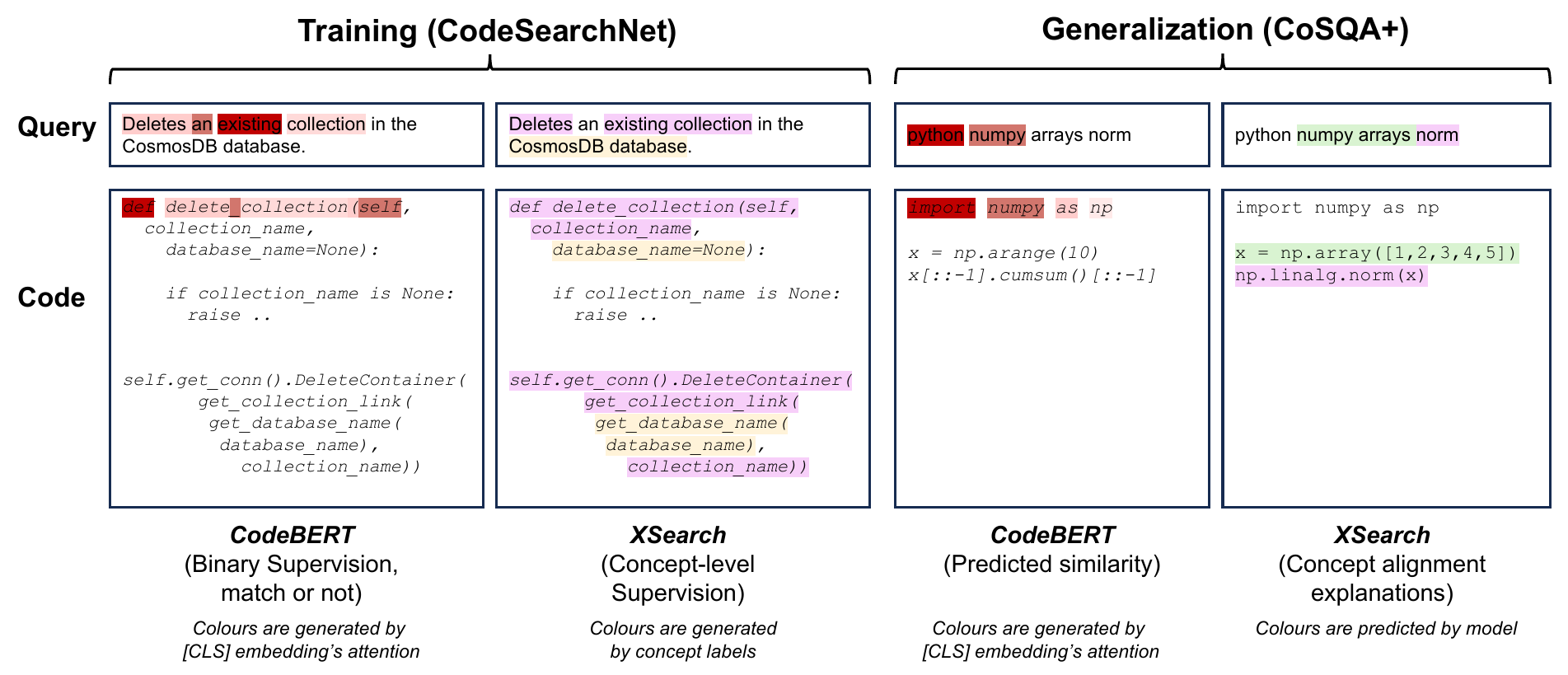}
  \caption{
  A CodeBERT retriever trained on CodeSearchNet exhibits shortcut learning, relying on leading tokens (e.g., function names) and failing on CoSQA+, where key semantics appear elsewhere.
  In contrast, \ourmethod retrieves by concept-to-code alignment, making the decision traceable and more robust.
  (Colours: CodeBERT uses post-hoc [CLS] attention.
   \ourmethod shows predicted concept highlights and alignments.)
  }\label{fig:motivating-example}
\end{figure*}

\textbf{C1: The Interpretability Gap (Post-hoc v.s. Intrinsic).}
Developers may encounter retrieved code that appears lexically similar yet fails to satisfy some critical requirements of the query, 
while the model provides no rationale for why the result was retrieved.
Given a query–code pair, developers are presented with only a similarity score, without knowing \textit{which part of the code satisfies which part of the query}.
This lack of transparency makes it difficult to decide whether to accept or reject the recommended code.
While post-hoc explanation techniques (e.g., Attention maps, XCos \cite{wang2023xcos}, EXS \cite{singh2019exs}) exist, they follow a "predict-then-explain" paradigm.
Code is first retrieved using a standard retriever, and explanations are generated afterwards using external mechanisms such as knowledge graphs.
As a result, the explanation does not influence how representations are learned and how retrieval decisions are made.
To address this, \ourmethod unifies the prediction and explanation processes. In our approach, the concept-to-code alignments \textit{are} the explanations, ensuring that the rationale is intrinsic to the model's decision-making rather than generated post-hoc.

\textbf{C2: The Generalization Collapse.}
Existing retrievers often fail to generalize when the test data distribution differs from the training data~\cite{hendrycks2020pretrained, chen2024decoder}.
As illustrated in Figure~\ref{fig:motivating-example}, models trained on CodeSearchNet tend to rely on ``low-hanging'' cues such as leading tokens (e.g., function name) to match queries with code.
While such cues are highly predictive in-distribution, they become unreliable when code structure or naming conventions change.
Consequently, retrieval performance collapses in out-of-distribution settings.
Recent studies further suggest that increasing model scale to billions of parameters can partially improve robustness, but does not fundamentally eliminate shortcut reliance~\cite{chen2024decoder}.
To overcome this, \ourmethod shifts from using global \texttt{[CLS]} similarity to per-concept matching. By explicitly requiring that \textit{every} query concept be grounded in specific code statements, our model structurally prevents reliance on partial token overlaps and superficial shortcuts.

\textbf{Root Cause: Inductive Matching vs. Deductive Validation.}
We argue that both challenges stem from a fundamental inductive bias
in existing retrievers. 
Current models are trained to \emph{inductively learn} 
statistical correlations between queries and code via global embedding matching with binary supervision. 
However, developers assess code \emph{deductively} by 
verifying whether each functional requirement in the query is satisfied. 
This mismatch explains why scaling model capacity alone cannot eliminate 
shortcut learning: the problem lies not in model size, but in the learning paradigm itself.

Motivated by this observation, we propose \ourmethod, which shifts code 
search from inductive matching to deductive alignment. 
Concretely, we reformulate retrieval as a \textit{concept-to-code alignment 
problem}, where the model must explicitly verify that each query concept 
is grounded in the candidate code.
Our key insight is that code search queries are often \emph{compositional}, consisting of multiple functional concepts (e.g., actions, entities, and constraints).
The correct code typically implements these concepts in certain localized code regions.
Instead of collapsing the entire query and code into single vectors, 
\ourmethod
(i) first decomposes a query into concepts, 
(ii) identifies salient code spans that implement similar functionality, and (iii) then computes an optimal concept-to-code alignment to rank candidates and generate explanations.
Importantly, concept alignment is integrated into both training and retrieval, encouraging the model to cover multiple functional aspects of the query rather than relying on global similarity alone.
This design produces concept-level explanations that can help developers assess whether retrieved code satisfies their functional intent.

We evaluate \ourmethod against six state-of-the-art encoder-only retrievers (CodeBERT~\cite{feng2020codebert}, UniXCoder~\cite{guo2022unixcoder}, GraphCodeBERT~\cite{guo2020graphcodebert}, CoCoSoDa~\cite{shi2023cocosoda}, HedgeCode~\cite{chen2024hedgecode}, RAPID~\cite{fan2024rapid}) as well as two decoder-only retrievers (Qwen2.5-Coder-7B-lt-SupCon-CSN and CodeLlama-lt-SupCon-CSN \cite{chen2024decoder}).
Our model is fine-tuned on the GraphCodeBERT encoder with approximately 125M parameters.
On out-of-distribution benchmarks, existing retrievers struggle significantly, even for decoder models with 7B parameters. 
\ourmethod effectively bridges this gap, improving performance from negligible levels to practical usability (e.g., boosting MRR from $0.02$ to $0.33$).
In addition, a controlled user study demonstrates that concept-alignment explanations enable participants to accept or reject retrieved results 38\% faster and 10.5\% more accurately than when only similarity scores are provided.

We summarize the contributions of this paper as follows:
\begin{itemize}[leftmargin=*]
    \item  
    We identify a fundamental limitation of existing code retrievers: 
    their inductive formulation encourages shortcut learning. 
    We address this by reformulating 
    code search as deductive concept alignment, enabling the model 
    to verify each functional requirement explicitly.
    
    \item 
    We propose \ourmethod, an intrinsically explainable code retrieval framework, 
    where concept-level alignments are unified into the model's prediction,  rather than generated in a post-hoc way.

    \item 
    We develop an LLM-assisted label augmentation pipeline to construct concept annotations for training an alignment-aware retrieval model.
    We open-source both the label augmentation framework and the dataset in \cite{xsearch}.
    
    \item 
    We extensively evaluate \ourmethod with eight state-of-the-art code retrievers on seven code search tasks.
    In addition, we design a user study consisting of 20 participants over 20 code search tasks.
    The results demonstrate great improvements in both explainability and performance generalizability.
\end{itemize}


%% file: motivating_example.tex


\section{Preliminary}
In this section, we first introduce related work on code search, then we present our motivating example.

\subsection{Background}

\noindent\textbf{LLMs for Code.}
In recent years, large language models (LLMs) have been widely adopted for code-related tasks. Some serve as foundational models pre-trained on general code-solving tasks, while others are specialized for specific domains.
General-purpose code LLMs are pre-trained on typical NLP tasks such as masked language modeling \cite{codesage, codet5, codet5+, guo2022unixcoder}, span denoising \cite{codesage, codet5+, guo2022unixcoder},  next-sequence prediction \cite{deepseek-coder, guo2022unixcoder}, and fill-in-the-middle prediction \cite{startcoder, deepseek-coder}. 
Some research works incorporate code-specific tasks such as AST edge prediction \cite{syncobert}, data flow edge prediction \cite{guo2020graphcodebert}, identifier tagging \cite{syncobert,codet5}, and code-NL contrastive learning \cite{syncobert, codesage, codet5, codet5+, guo2022unixcoder}.

Meanwhile, specialized LLMs are designed to address specific tasks.
The tasks can be grouped into the following categories:
1) NL-to-code generation: code generation from documentation \cite{huang2023enhancing, chen2024jumpcoder, li2025programming, dai2024mpcoder, bostrom2024language},
2) NL-to-code retrieval: code search \cite{yu2024droidcoder, ni2023lever, li2022coderetriever, inala2022fault, chen2024hedgecode, shi2023cocosoda}.
3) Code-to-NL generation: code summarization \cite{ahmad2020transformer, tang2022ast, shi2023coss, su2024distilled} and code-to-comment generation \cite{kang2024identifying, bappon2024autogenics, li2022auger}.
4) Code-to-code generation: code completion \cite{jiang2024aixcoder, zhang2023repocoder, liu2024graphcoder}, code repair \cite{yasunaga2021break, fan2023automated, fu2022vulrepair, peng2024domain, huq2022review4repair} and code translation \cite{chen2018tree, lachaux2020unsupervised, roziere2021leveraging, szafraniec2022code, huang2023program, tehranijamsaz2024coderosetta}.
Beyond these applications, LLMs are increasingly used to streamline the software development process, including tasks such as commit message generation \cite{xue2024automated, li2025optimization, jung2021commitbert, eliseeva2023commit}, automated code review \cite{tufano2022using, tang2024codeagent}, refactoring suggestions \cite{stengel2024regal}, and detecting security vulnerabilities \cite{li2018vuldeepecker, li2021sysevr, hellendoorn2019global, nong2022open, steenhoek2023empirical, wang2024llmdfa, ji2024applying}.

\noindent\textbf{Code Search.}
In this work, we focus on the code search task.
Code search aims to retrieve relevant code snippets given a natural-language query \cite{di2023code, tang2023hyperbolic, kim2018facoy,liu2021opportunities,chen2024code}.
It can be integrated into downstream applications such as code generation \cite{chen2024code}, fault localization \cite{jiang2025cosil}, and program repair \cite{zhang2024no}.
Existing approaches can be broadly categorized into \textit{keyword-based} methods and \textit{embedding-based} semantic methods.
\begin{itemize}[leftmargin=*]
    \item \textbf{Keyword-based code search.}
Early code search systems relied on lexical matching between a natural-language query and code artifacts, using information-retrieval techniques over identifiers, comments, and API names~\cite{chan2012searching,mcmillan2011portfolio,lv2015codehow}. To reduce vocabulary mismatch, follow-up work explored query expansion and reformulation, e.g., by adding synonyms or related terms to the query~\cite{nie2016query,hill2014nl,lu2015query,raghothaman2016swim}. 
While efficient and transparent, keyword-based retrievers are limited by surface-form overlap and often fail when relevant code uses different wording, naming conventions, or abstractions.
\item \textbf{Embedding-based semantic code search.}
To overcome lexical limitations, embedding-based approaches map queries and code snippets into a shared embedding space, where relevance is measured by vector similarity.
Early neural network approaches such as Deep Code Search \cite{gu2018deep} and CodeSearchNet \cite{husain2019codesearchnet} learn joint representations for code and natural language.
Recent Transformer-based models, including CodeBERT \cite{feng2020codebert}, GraphCodeBERT \cite{guo2020graphcodebert}, UniXcoder \cite{guo2022unixcoder}, and CodeRetriever \cite{li2022coderetriever}, leverage large-scale pretraining and self-attention to further improve retrieval accuracy.
More recent methods such as CoCoSoDa \cite{shi2023cocosoda} and HedgeCode \cite{chen2024hedgecode} adopt contrastive learning objectives to strengthen representation alignment.
Besides, recent efforts are being made to improve accuracy by deepening search steps \cite{gu2025spencer}, improving searching efficiency \cite{gu2022accelerating}, and enhancing data augmentation strategies \cite{li2022exploring}.
\end{itemize}

\subsection{Motivating Example}\label{sec:motivation-eg}

\autoref{fig:motivating-example} shows an observed shortcut learning behavior for an encoder-based code retriever trained with CodeBERT \cite{feng2020codebert} on the CodeSearchNet benchmark \cite{husain2019codesearchnet}.
The figure shows four query-code pairs.
The first pair is a training example from CodeSearchNet with a binary label (i.e., whether they are relevant or not).
The second pair is the same training example with our proposed concept-level label augmentation, where query concepts are explicitly aligned with code units.
The third pair is a top-ranked result retrieved by CodeBERT, on out-of-distribution benchmark CoSQA+ \cite{gong2024cosqa+}.
And the fourth pair is the top-ranked result retrieved by \ourmethod.
In addition, we highlight the focal tokens that contribute the most to the model's prediction.
For CodeBERT, contributions are measured by post-hoc methods, and we trace the attention weights by using \texttt{[CLS]} embedding as query and all tokens as keys.
For \ourmethod, we visualize the predicted concept-bearing tokens and their concept alignments directly.

\noindent\textbf{Observation: Shortcut Learning in CodeBERT.}
On CodeSearchNet, CodeBERT assigns high contribution to leading tokens such as function names.
This behavior is effective in-distribution (first pair), since many code sample conveys their functionality in function names. 
However, when applied to out-of-distribution data where such naming conventions do not hold, the model relies on the same positional cues and retrieves incorrect results, as shown in the third pair.

\noindent\textbf{Key Difference with Concept-level Supervision.}
In contrast, \ourmethod explicitly models retrieval as a concept alignment problem.
Instead of relying on global embedding similarity, it aligns each functional concept in the query with corresponding code units.
As shown in the fourth pair, the retrieve code covers all the required concepts ``numpy arrays'' and ``norm''.
This reduces the model's tendency to over-reliance on positional or naming cues.

\noindent\textbf{Our Solution: Label Augmentation.}
Different from the traditional data augmentation to improve the model generalizability, we adopt a novel label augmentation in this work.
Specifically, we annotate the training samples by
(1) how to extract the concept in the query and
(2) how those concepts are present and absent in the code.
This mirrors how developers assess code relevance in practice, where missing any required operation often renders a candidate unusable.
By training the model to reason over concept coverage instead of surface patterns, \ourmethod learns representations that are both more interpretable and more robust to distribution shift.

%% file: approach.tex
\begin{figure*}[t]
    \centering
    \includegraphics[width=\linewidth]{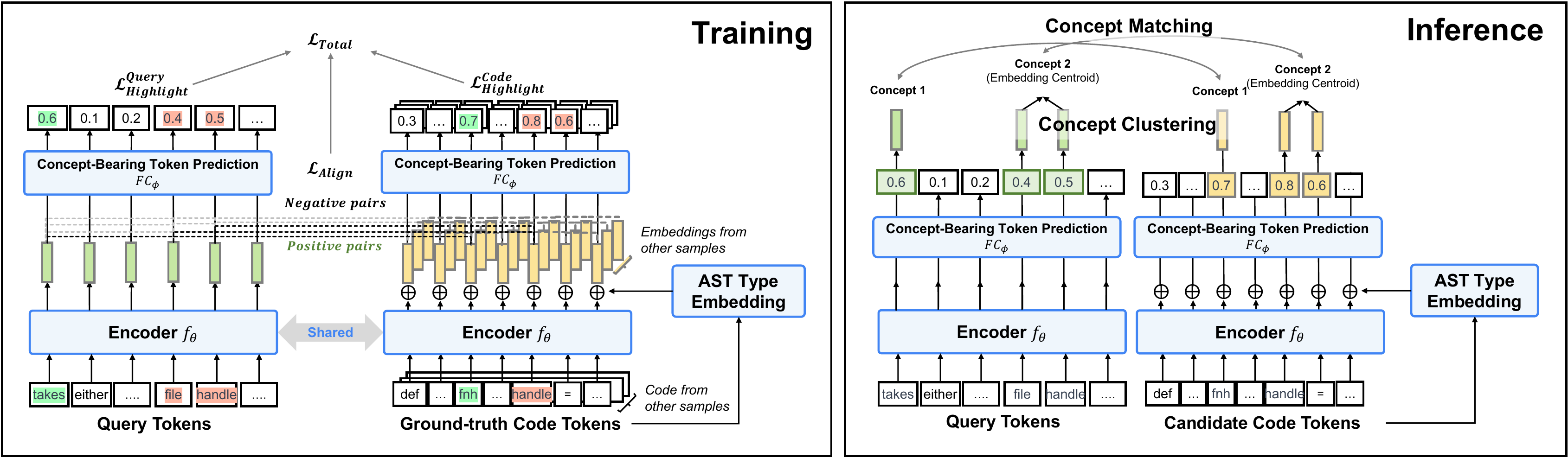}
    \caption{\ourmethod Model Architecture.
    During training (LHS), concept-aligned query and code token spans form positive pairs, while non-aligned spans form negative pairs. 
    A shared encoder maps query and code tokens to contextual embeddings.
    For code tokens, AST type embeddings are added to incorporate structural information. 
    A linear probing head predicts concept-bearing tokens at the token level, and an alignment objective trains the model to associate query concepts with their corresponding code spans.
    At inference time (RHS), the model identifies concept-bearing tokens in both query and code, aggregates them into concept representations, and matches query concepts to code concepts based on embedding similarity.
    }
    \label{fig:model}
\end{figure*}

\section{Approach}\label{sec:approach}



\subsection{Overview}\label{sec:app-overview}

The input is a natural-language query $n$ and the goal is to retrieve code snippets $c \in \mathcal{C}$ that best satisfy the query. 
We propose a novel retrieval model architecture, as shown in Figure~\ref{fig:model}.
In addition to retrieving relevant code, \ourmethod also produces explanations by explicitly aligning code lines with the concepts mentioned in the query. 
We develop \ourmethod through three stages:
\begin{itemize}[leftmargin=*]
    \item \textbf{Stage 1: Concept Label Augmentation (Section \ref{sec:annot}).}
    We annotate essential concepts in queries and identify the corresponding code units that implement each concept. 
    This stage produces token-level query-code alignment annotations.
    
    \item \textbf{Stage 2: Alignment-Aware Model Training (Section \ref{sec:model-train}).}
    Using the annotated alignments, we train an alignment-aware retrieval model that jointly identifies concept-bearing tokens and aligns semantically related query-code pairs in the embedding space.
    
    \item \textbf{Stage 3: Explainable Query-to-Code Retrieval (Section \ref{sec:retrieval}).}
    At inference time, the model aligns predicted concept-bearing tokens between the query and candidate code snippets, retrieves the top code by similarity, and outputs the corresponding concept alignment maps as explanations.
\end{itemize}

\input{approach/label_augmentation}
\input{approach/train}

\input{approach/retrieval}

%% file: approach/label_augmentation.tex
\subsection{Concept Label Augmentation}\label{sec:annot}

We perform \emph{label augmentation} rather than data augmentation.
For each existing query-code pair, we annotate 
(i) concept spans in the query and
(ii) alignments from each concept to the code units that implement it.
This produces fine-grained supervision without creating new training examples.
Because such annotations must be both scalable and reliable, we combine LLM-based annotation with strict validation.

\subsubsection{Concept Definition.}
We define a concept as a set of query tokens 
that together express a single, indivisible functional requirement.
Internally, a concept is compositional:
it typically consists of an \textbf{Action} or \textbf{Entity}, and optionally combined with \textbf{Modifiers} that specify constraints or attributes 
(Table~\ref{tab:def-concept}).
For example, ``delete existing collection'' forms a unified concept with an action ``delete'', an entity ``collection'', and a modifier ``existing''.
Each query token is assigned to at most one concept (or marked as non-concept).
We do not assume concepts to be perfectly defined.
Rather, we only require that constraint-bearing units can be identified with reasonable consistency under our definition.
Specifically, a token is considered ``concept-bearing'' if it belongs to a concept expressing a single functional requirement. For example, in the query ``delete an existing collection'', the tokens ``delete'', ``existing'', and ``collection'' are concept-bearing, whereas ``an'' is not. On the code side, the corresponding aligned unit consists of the lines that perform the deletion (e.g., \texttt{self.get\_conn().DeleteContainer}). Tokens from those specific lines are considered concept-bearing code tokens.

\begin{table}[t]
    \centering
    \scriptsize
    \caption{Concept Components and Examples.}\label{tab:def-concept}
    \begin{tabular}{p{3cm} p{5.5cm}}
    \toprule
    \textbf{Component} & \textbf{Description} \\
    \midrule
    Action & Operation to perform (\textit{delete}, \textit{return}) \\
    Entity & Target object (\textit{collection}, \textit{file handle}) \\
    Modifier & Constraint/attribute (\textit{existing}, \textit{open}) \\
    \midrule
    \textbf{Concept Example} & \textbf{Composition} \\
    \midrule
    \textit{delete existing collection} & Action + Modifier + Entity \\
    \textit{open file handle} & Modifier + Entity \\
    \bottomrule
    \end{tabular}
\end{table}

\begin{figure*}[t]
    \centering
    \includegraphics[width=\linewidth]{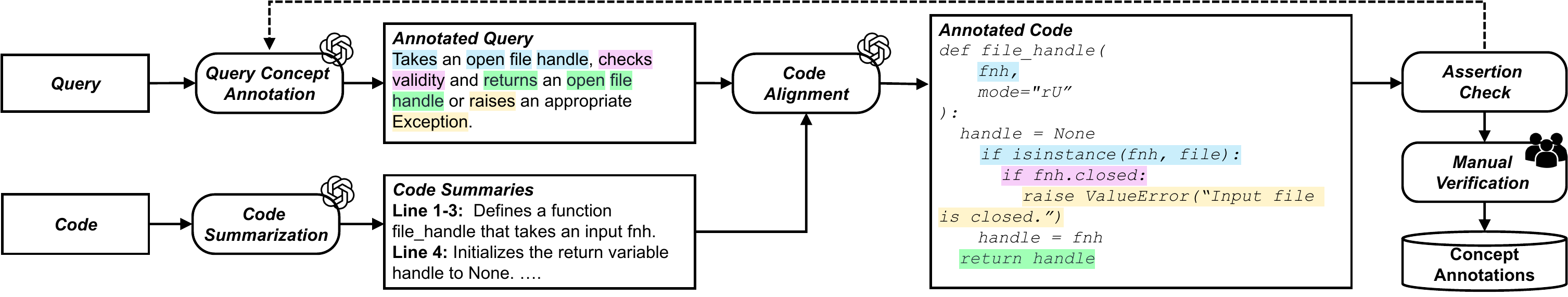}
    \caption{LLM-assisted Annotation Pipeline.}
    \label{fig:annotation-pipeline}
\end{figure*}

\subsubsection{LLM-Assisted Annotation Pipeline.}\label{sec:llm-oracle}
Given a query $n$ and its paired code snippet $c$ in the dataset, we annotate them in three steps (Figure~\ref{fig:annotation-pipeline}).

\begin{enumerate}[leftmargin=*]
\item \textbf{Step 1: Query Concept Decomposition.}
We ask the LLM to identify the concepts in the query.
A concept $a$ is represented as a set of query tokens
$\mathcal{N}_a = \{n_i \mid \phi(n_i)=a\}$, where $\phi(\cdot)$ assigns each token to a concept or to \textit{None}.

\item \textbf{Step 2: Code Summarization.}
We ask the LLM to split the code into code units (by default, code lines, consecutive lines can be merged when they form one coherent statement),
and ask it to write a short natural-language description $d$ for each unit.
This provides an intermediate textual description that is easier to align with query concepts.

\item \textbf{Step 3: Code Alignment.}
We align each query concept $a$ to one or more code units that implement it.
We then map aligned units back to code tokens, producing
$\mathcal{C}_a = \{c_j \mid \phi(c_j)=a\}$.
Unlike queries, a code unit (and its tokens) can be linked to multiple concepts.
\end{enumerate}

\subsubsection{Assertion-Based Validation.}\label{sec:assertion}
To improve annotation reliability, we validate each LLM output using two types of assertions:
(1) \textbf{format assertions} to ensure that the output can be parsed, and
(2) \textbf{consistency assertions} to ensure that every concept span and aligned code unit \textcolor{blue}{appears verbatim} in the original query and code. Any annotation that introduces non-existing tokens (i.e., hallucination) is strictly rejected.
If any assertion fails, we return the error messages to the LLM and retry from Step~1.
We discard samples that fail after $R$ retries.
In our data construction, 72.98\% pass with no retry, 97.11\% pass within two retries, and all accepted samples pass within five retries.

\subsubsection{Small-Scale Manual Verification.}\label{sec:manual-verify}
To assess annotation quality, we randomly sampled 1,000 annotated pairs per programming language and asked two developers (with 3 years+ programming experience) to independently judge whether
(i) query concept spans follow the definition in Table~\ref{tab:def-concept}, and
(ii) aligned code units reasonably correspond to the intended concepts.
The inter-annotator agreement is high ($\kappa=0.9$), 
suggesting that the annotations are largely consistent under our definition, although we do not claim this procedure fully eliminates semantic errors.
Furthermore, we found that in 88.3\% of the cases, human annotators agreed with the LLM-annotated alignments and concept spans.

\input{tables/assertion}

%% file: approach/train.tex
\subsection{Alignment-Aware Model Training}\label{sec:model-train}

To jointly learn (i) concept-bearing tokens and (ii) concept-to-code alignments, we optimize two objectives: a token-level \textbf{highlight loss} and a span-level \textbf{alignment loss}.
The highlight loss predicts whether each token should be highlighted as concept-bearing.
The alignment loss pulls each query concept span close to its aligned code span while pushing it away from hard negatives.

\subsubsection{\textbf{Highlight Loss}}\label{sec:highlight-loss}

We add a lightweight probing head on top of the encoder to predict a highlighting probability for each token.
Given an input sequence $x$ (a query or code snippet), the encoder $f_{\theta}$ outputs contextual hidden states ${h}_{i=1}^{N}$, where $h_i=f_{\theta}(x_i)$ is the representation of token $x_i$.
A linear classifier produces $p_i\in(0,1)$:
\begin{equation}
p_i = \sigma(\mathrm{FC}_{\phi}(h_i)) = \sigma(\mathrm{FC}_{\phi}(f_\theta(x_i))) ,\quad x_i \in {\{\text{query tokens, code tokens}\}}.
\end{equation}
Since concept-bearing tokens are sparse, we use focal loss to address the class imbalance problem \cite{lin2017focal}:

\begin{equation}\label{eq:highlight-loss}
\begin{split}
\mathcal{L}_{\mathrm{high}}
&= - \sum_{i=1}^{N} 
   \Bigl[
     \alpha\,(1 - p_i)^{\gamma}\,y_i\,\log p_i
     \;+\;
     (1 - \alpha)\,p_i^{\gamma}\,(1 - y_i)\,\log(1 - p_i)
   \Bigr],
\end{split}
\end{equation}
where $y_i \in \{0,1\}$ indicates whether token $x_i$ is concept-bearing, and $\alpha,\gamma$ are focal-loss hyperparameters.
Note that when $\gamma = 0$ and $\alpha = 0.5$, the focal loss reduces to standard binary cross-entropy.

For code tokens, we additionally incorporate coarse-grained AST node types (20 types) via a type embedding layer.
Type embeddings are added to token representations before the probing head.
We use separate probing heads for queries and code.
%


\subsubsection{\textbf{Alignment Loss with Hard Negative Sampling}}\label{sec:alignment-loss}

Let a query concept span be $\mathcal{N}_a$ (a set of query tokens belonging to concept $a$), and its aligned code span be $\mathcal{C}_a$ (a set of code tokens aligned to the same concept).
We represent a span by mean-pooling its token embeddings:
\begin{equation}
f(\mathcal{S}) = \frac{1}{|\mathcal{S}|}\sum_{x_i \in \mathcal{S}} f_\theta(x_i),\quad \mathcal{S} = \mathcal{N}_a \text{ or } \mathcal{C}_a.
\end{equation}
We then apply a contrastive objective: for each $\mathcal{N}_a$, treat $(\mathcal{N}_a,\mathcal{C}_a)$ as a positive pair and sample $K$ hard negatives $\mathcal{S}_{a,[K]}^{-}$ from
(i) \emph{intra-sample negatives} (other concepts in the same query-code pair) and
(ii) \emph{inter-sample negatives} (concept spans from other pairs in the batch).
The full algorithm is presented in Algorithm~\ref{alg:truf_neg_sampling}.

Specifically, to prioritize ``hard'' negatives that are spuriously similar under the current model, we score each negative $\mathcal{C}_b$ ($b\neq a$) by:
\begin{equation}\label{eq}
\mathrm{score}_{b} \;=\;
\underbrace{\cos \; \bigl(f_{\theta}(\mathcal{N}_a), f_{\theta}(\mathcal{C}_b)\bigr)}_{\text{current model}}
\; - \;
\underbrace{\cos \; \bigl(g(\mathcal{N}_a), g(d_b)\bigr)}_{\text{reference text model}},
\end{equation}
where $f_{\theta}(\cdot)$ uses our encoder representations, 
and $g(\cdot)$ is a frozen sentence encoder (all-mpnet-base-v2 \cite{song2020mpnet}) 
pretrained on semantic similarity tasks.
The reference model is used \emph{only} to estimate negative hardness and is independent of our model.
Its input is $\mathcal{N}_a$ and $d_b$, the natural-language description of the code span $\mathcal{C}_b$ (from Step~2 in the annotation pipeline described in Section~\ref{sec:annot}).
Intuitively, a high score indicates that the current model considers $(\mathcal{N}_a,\mathcal{C}_b)$ similar, while the reference text model considers $\mathcal{N}_a$ and $d_b$ semantically mismatched.
In other words, we prioritize negatives that the current model 
mistakenly finds similar, these are the most informative for training.

\begin{algorithm}[h]
\small
\caption{Hard Negative Sampling for Alignment Loss}
\label{alg:truf_neg_sampling}
\begin{algorithmic}[1]
  \REQUIRE A batch of query–code pairs; Current model $f_\theta$, reference model $f_{\mathrm{ref}}$; Number of negatives per concept as $K$  
  \ENSURE Query concepts and their Top K hard negatives:
  $\{(\mathcal{N}_a, \mathcal{S}_{a, [K]}^-)\}$
  \FOR{query concept $\mathcal{N}_a$}
    \STATE $\mathcal{S}_a^-
        \gets \{\, \mathcal{C}_b \mid  b \neq a\}$ // Collect all negative code spans
    \FOR{code span $\mathcal{C}_b \in \mathcal{S}_a^-$}
    \STATE $\mathrm{score}_b \gets 
      \cos \; \bigl(f_{\theta}(\mathcal{N}_a), f_{\theta}(\mathcal{C}_b)\bigr)
        \; - \;
        \cos \; \bigl(g(\mathcal{N}_a), g(d_b)\bigr)$
    \ENDFOR
    \STATE // Select Top $K$ hardest negatives
    \STATE Sort $\mathcal{S}_a^-$ by $\mathrm{score}_b$ in descending order
    \STATE $\mathcal{S}_{a, [K]}^- \gets \text{first $K$ elements of }\mathcal{S}_a^-$
  \ENDFOR
  \RETURN $\{(\mathcal{N}_a, \mathcal{S}_{a, [K]}^-)\}$
\end{algorithmic}
\end{algorithm}

After negative sampling, we employ the contrastive learning loss.
We adopt InfoNCE loss \cite{infonce} as follows.

\begin{equation}\label{eq:align-instance}
\begin{split}
\mathcal{L}_{\text{Align}} &= 
\sum_{\mathcal{N}_a} -\log
  \frac{
    \exp\bigl(u_a/\tau\bigr)
  }{
    \exp\bigl(u_a/\tau\bigr)
    + \sum\limits_{\mathcal{C}_b \in \mathcal{S}_{a, [K]}^-}
       \exp\bigl(v_{a,b}/\tau\bigr)
  },\\
\text{where}\quad
u_a &= \cos\bigl(f_\theta(\mathcal{N}_a),f_\theta(\mathcal{C}_a)\bigr),
v_{a,b} = \cos\bigl(f_\theta(\mathcal{N}_a),f_\theta(\mathcal{C}_b)\bigr). 
\end{split}
\end{equation}

Here, $\mathcal{N}_a$ is the a-th query‐concept span's embedding, 
$\mathcal{C}_a$ is its positive code‐span embedding,
$\mathcal{S}_{a, [K]}^-$ is the set of $K$ hardest negatives for $\mathcal{N}_a$,
and $\tau$ is the temperature hyper-parameter.
The overall training objective becomes:
\begin{equation} 
\mathcal{L}_{\text{Total}} = \mathcal{L}_{\text{High}}^{\text{Code}} 
+ \mathcal{L}_{\text{High}}^{\text{query}} + \mathcal{L}_{\text{Align}} 
\end{equation}
We use equal weights as the three losses are empirically on comparable scales.

%% file: approach/retrieval.tex
\subsection{Explainable Query-to-Code Retrieval}\label{sec:retrieval}

At inference time, the input consists of a natural-language query $N_{\text{query}}$ and a reference codebase ${C_{\text{ref}}}$. 

\paragraph{\textbf{Concept Clustering for Query.}}
Given a query, we run inference with the retrieval model and obtain the highlight score $p_i$ for token $i$. 
Tokens with $p_i > \delta_{\text{highlight}}$ are selected as concept-bearing and grouped using agglomerative hierarchical clustering with cosine similarity.
Clusters whose centroid similarity exceeds a threshold $\delta_{\text{cluster}}$ are merged. 
Each resulting cluster $C$ is represented by a centroid
\begin{equation}\label{eq:query-concept-cluster}
    \mathbf{h}_C = \frac{1}{|C|} \sum_{n_i \in C} f_\theta(n_i),
\end{equation}
which serves as a concept-level representation of the query. 
We set $\delta_{\text{highlight}} = 0.4$ and $\delta_{\text{cluster}} = 0.8$
(sensitivity analysis in Figure~\ref{fig:hyperparam_sweep}).

\paragraph{\textbf{Concept Matching for Code.}}
For code, we adopt a line-level representation rather than clustering.
Unlike queries where concept boundaries must be inferred, 
code has explicit structural boundaries: 
each line typically corresponds to a single statement.
For each non-empty code line $l$, we check whether it contains at least one 
concept-bearing token (i.e., $\exists\, c_j \in l$ such that $p_j > \delta_{\text{highlight}}$).
Lines without any concept-bearing tokens are excluded from matching.
For retained lines, let $\mathcal{C}_l$ denote \emph{all} tokens in line $l$, 
and we compute a line-level embedding:
\begin{equation}\label{eq:code-line}
    \mathbf{e}_l = \frac{1}{|\mathcal{C}_{l}|}\sum_{c_j \in \mathcal{C}_{l}}{f_\theta(c_j)}.
\end{equation}
Given query-concept centroids $\{\mathbf{h}_{C_k}\}_{k=1}^{K}$ and 
code-line embeddings $\{\mathbf{e}_l\}_{l=1}^{L}$, 
we apply greedy matching: each query concept is matched to the code line 
with the highest cosine similarity.
The overall similarity is the average across all matches:
\begin{equation}
    \text{sim}(q, c) = \frac{1}{K} \sum_{k=1}^{K} \max_{l} \cos(\mathbf{h}_{C_k}, \mathbf{e}_l).
\end{equation}
This ensures that the retrieved code must address all query concepts to score highly.
Note that \ourmethod is designed to be conservative at inference time, prioritizing constraint satisfaction over partial matches.

\subsubsection{Output Explanation Format}
Figure~\ref{fig:ui_eg} shows an example output.
Given the query, \ourmethod identifies two concepts and returns the top-2 code snippets.
For each snippet, it highlights the lines that support each concept.
The top-2 results both implement dictionary merging and therefore receive high similarity.
Code B is closer because it satisfies the requirement on ``overwriting existing keys'', while Code A doesn't.
Overall, \ourmethod not only ranks code snippets by semantic relevance, but also provides explicit concept-to-code alignments, reducing developers' cognitive effort when deciding whether to accept or reject a recommendation.
We evaluate this benefit in our user study (Section~\ref{sec:user-study}).

\subsubsection{Complexity Analysis}
We analyze the inference complexity of \ourmethod.
Let $a$ and $b$ denote the token lengths of the query and code, respectively.
The encoder forward pass costs $\mathcal{O}(a)$ and $\mathcal{O}(b)$.
Let $k$ be the number of highlighted tokens in the query ($k < a$).
Query concept clustering costs $\mathcal{O}(k^2)$, while centroid computation is $\mathcal{O}(k)$.
For code, computing line-level embeddings (Equation~\ref{eq:code-line}) costs $\mathcal{O}(b)$, and 
concept matching costs $\mathcal{O}(KL)$, where $K$ is the number of query concepts 
and $L$ is the number of code lines.
Overall, the inference complexity is
$\mathcal{O}\left(a + k^2 + |C_{\text{ref}}| \cdot (b + KL)\right)$.
Since $k$, $K$, and $L$ are small in practice, the complexity is dominated 
by encoder computation, comparable to a standard bi-encoder.

%% file: experiments/main.tex
\section{Experiments}

\subsection{Hyperparameter Setup}
\begin{figure}[t]
    \centering
    \includegraphics[width=0.7\linewidth]{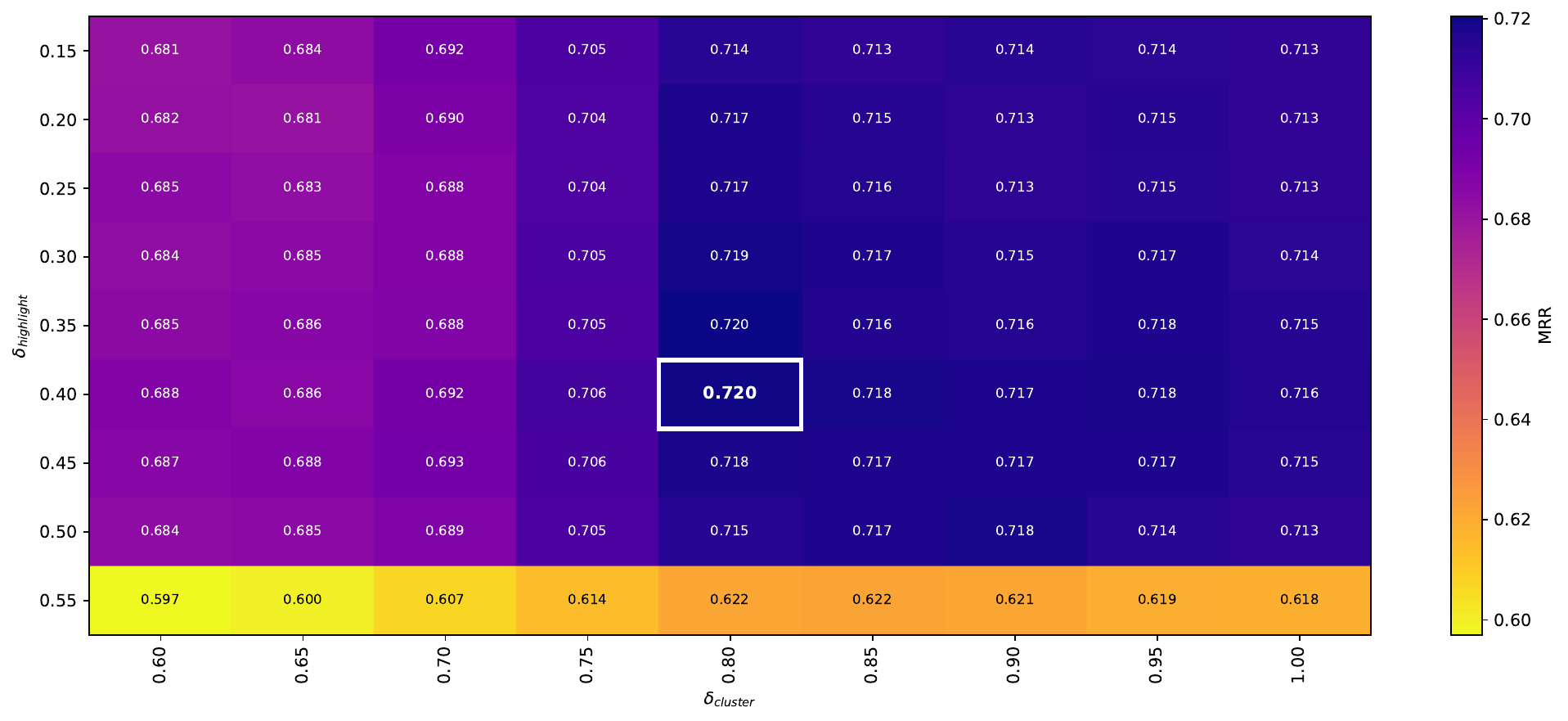}
    \caption{
    Hyperparameter sweep on the CodeSearchNet-python \cite{husain2019codesearchnet} test set. 
    We vary $\delta_{highlight}$ and $\delta_{cluster}$ within a focused range and report MRR. Darker color indicates higher MRR.}
    \label{fig:hyperparam_sweep}
\end{figure}

In the training stage, we fine-tune GraphCodeBERT \cite{graphcodebert} on CodeSearchNet \cite{husain2019codesearchnet}, use a batch size of 32, an epoch number of 10, Adam optimizer with initial learning rate $2e-5$, and linear learning rate decay.
We set $\gamma = 2.0, \alpha = 0.5$ for highlight loss.
For alignment loss, we use the reference model as ``all-mpnet-base-v2'' \cite{song2020mpnet}, which is pretrained on natural language sentence pairs.
We sample the top-50 ranked negative code for alignment loss computation, and we set temperature $\tau = 0.1$ (same as \cite{guo2020graphcodebert}).
At the inference stage, we first identify concept-bearing tokens by thresholding the highlight scores with $\delta_{highlight}$ and then merge the retained tokens into concepts using the cluster threshold $\delta_{cluster}$. Based on a small-scale experiment on the test set (\autoref{fig:hyperparam_sweep}), we set $(\delta_{highlight}, \delta_{cluster})=(0.4, 0.8)$ and find the performance to be stable under modest variations of the thresholds.

\subsection{Research Questions}
To evaluate our approach, we conduct extensive experiments designed to answer four key research questions (RQs): 

\noindent\textbf{RQ1: Overall Retrieval Performance:} 
How does \ourmethod compare with existing methods, on both in-distribution and out-of-distribution code search benchmarks?


\noindent\textbf{RQ2: Faithfulness of Concept Explanations:} 
How accurate are the two explanation components produced by \ourmethod: 
(i) highlighted concept-bearing tokens and 
(ii) concept-to-code alignments?


\noindent\textbf{RQ3: Ablation Study:} 
How do the key components of \ourmethod (e.g., concept highlighting, negative sampling in alignment training) affect retrieval performance?


\noindent\textbf{RQ4: User Study:} 
Do concept-alignment explanations help developers judge retrieved results more quickly and more accurately than similarity scores alone?

Given the space limit, more experimental results, demonstrations, and qualitative analysis are available at \cite{xsearch}.




\input{experiments/rq1}
\input{experiments/rq2}
\input{experiments/rq3}
\input{experiments/rq4}

%% file: experiments/rq1.tex
\subsection{\textbf{RQ1: Overall Retrieval Performance}}

\subsubsection{\textbf{Setup}}

We evaluate \ourmethod on both in-distribution (ID) and out-of-distribution (OOD) benchmarks.
CodeSearchNet (CSN)~\cite{husain2019codesearchnet} is used for ID evaluation, providing large-scale one-to-one query–code pairs across six languages.
CoSQA+~\cite{gong2024cosqa+} is used for OOD evaluation, where each query is paired with multiple relevant code snippets, reflecting more realistic retrieval settings.
All models are fine-tuned on CSN and evaluated on CSN test sets for ID performance, and directly transferred to CoSQA+ for OOD evaluation.
Table~\ref{tab:datasets} reports dataset statistics.


\begin{table}[h]
    \centering
    \caption{Datasets Summary}
    \label{tab:datasets} 
    \resizebox{0.6\linewidth}{!}{
    \begin{tabular}{llrrrr}
        \toprule
        \textbf{Dataset} & \textbf{Language} & \textbf{\#Train} & \textbf{\#Valid} & \textbf{\#Test} & \textbf{\#Codebase} \\
        \midrule
        \multirow{6}{*}{\textbf{CodeSearchNet}}
        & Ruby       &  24,927  &  1,400  &  1,261   &  4,360  \\
        & JavaScript &  58,025  &  3,885  &  3,291   & 13,981  \\
        & Java       & 164,923  &  5,183  & 10,955   & 40,347  \\
        & Go         & 167,288  &  7,325  &  8,122   & 28,120  \\
        & PHP        & 241,241  & 12,982  & 14,014   & 52,660  \\
        & Python     & 251,820  & 13,914  & 14,918   & 43,827  \\
        \midrule
        \textbf{CoSQA$^+$} & Python & 16,481 & 2,058 & 2,065 & 51,516 \\
        \bottomrule
    \end{tabular}
    }
\end{table}

\input{tables/table_id_and_ood}
\par
\subsubsection{\textbf{Baselines}} 
We select the baselines to verify two key hypotheses:
(i) whether performance gains on out-of-distribution data can be achieved by introducing concept-to-code alignment objectives while keeping the same encoder architecture, 
and
(ii) whether such generalization can be achieved by scaling up model size alone, without reformulating the retrieval objective.

\paragraph{\textbf{Encoder-based Code Retrievers.}}
We include six widely used encoder-based retrievers.
All selected encoder baselines follow the bi-encoder retrieval paradigm, where queries and code are independently mapped into a shared embedding space.
CodeBERT \cite{feng2020codebert}, UniXcoder \cite{guo2022unixcoder}, and GraphCodeBERT \cite{guo2020graphcodebert} mainly enhance representations through pretraining strategies (e.g., masked language modeling, cross-modal, and multi-task pretraining).
CoCoSoDa \cite{shi2023cocosoda}, HedgeCode \cite{chen2024hedgecode}, and RAPID \cite{fan2024rapid} aim to improve robustness by modifying training objectives or data (e.g., contrastive learning, adversarial examples, or pseudo-labeled samples).
These models encode queries and code \texttt{[CLS]} token into a shared embedding space, and perform retrieval by their \texttt{[CLS]} similarity.
\ourmethod builds upon the GraphCodeBERT architecture and introduces an additional concept-bearing prediction head (Figure~\ref{fig:model}).

\paragraph{\textbf{Decoder-based Code Retrievers.}}
Recent work \cite{chen2024decoder} has shown that decoder-only LLMs achieve better generalization performance than encoder-only LMs when fine-tuned on code retrieval tasks.
Since decoder models do not have a special pooling token, they compute sequence-level embedding by taking the average embedding over all query/code tokens, and perform retrieval using vector similarity.
Following this setting, we include two strong LLM-based baselines released by  \cite{chen2024decoder}: \textit{Qwen2.5-Coder-7B-lt-SupCon-CSN} and \textit{CodeLlama-7B-lt-SupCon-CSN}.
They are fine-tuned from Qwen2.5-Coder-7B \cite{hui2024qwen2} and CodeLlama-7B \cite{roziere2023code} on CodeSearchNet, respectively.
These models represent the ``scale-up'' approach without changing the retrieval objective.


\subsubsection{\textbf{Evaluation Metrics}}
For each query, the retriever ranks code from the codebase by their relevance (the codebase size is detailed in Table~\ref{tab:datasets}).
We employ two metrics to evaluate retrieval performance.
\begin{itemize}[leftmargin=*]
\item \textbf{Mean Reciprocal Rank (MRR).} 
MRR is the standard metric for ranking-based retrieval tasks~\cite{graphcodebert, feng2020codebert}.
For each query, the reciprocal rank (RR) is defined as the inverse of the rank position of the first relevant result.
MRR is computed as the average RR over all queries:
\begin{equation}
\text{MRR} = \frac{1}{|Q|} \sum_{i=1}^{|Q|} \frac{1}{\text{rank}_i}
\end{equation}
where $|Q|$ is the number of queries,
and $\text{rank}_i$ is the rank position of the first relevant code snippet for the $i$-th query.

\item \textbf{Mean Multi-choice Reciprocal Rank (MMRR).} 
Unlike CodeSearchNet, queries in CoSQA$^+$ are associated with multiple relevant code snippets.
Following prior work, MMRR is used to evaluate CoSQA$^+$ \cite{chen2024decoder}.
It computes the highest rank among all correct answers:

\begin{equation}
\text{MMRR} = \frac{1}{|Q|} \sum_{i=1}^{|Q|} \max_{j \in R_i} \frac{1}{\text{rank}_{i,j}}
\end{equation}

where $R_i$ denotes the set of relevant code snippets for the $i$-th query, and $\text{rank}_{i,j}$ is the rank position of the $j$-th relevant snippet.
We adopt MMRR because it better reflects retrieval quality in multi-answer scenarios by evaluating whether a model can consistently rank any relevant result highly, 
rather than focusing on a single labeled match.
\end{itemize}

\subsubsection{\textbf{Results}}

Table~\ref{tab:generalization_performance} reports the retrieval performance of all models on both in-distribution (ID) CodeSearchNet and out-of-distribution (OOD) CoSQA+. 
On the in-distribution (ID) CodeSearchNet benchmark, encoder-only models show surprisingly high performance, often outperforming much larger 7B decoder models.
For example, CoCoSoDa achieves an MRR of 0.803 on the Ruby dataset.
However, their performances are near-zero on CoSQA+.
Decoder-based models obtain better transferability to OOD data (e.g., Qwen2.5-Coder-7B-lt-SupCon-CSN achieves an MRR of 0.24 on CoSQA+).
Nevertheless, these gains come at the cost of substantially larger training effort and higher inference overhead (Table~\ref{tab:runtime}).
\ourmethod achieves a better balance between efficiency and generalizability.
Its ID performance can be slightly lower when queries contain extra or overly specific constraints not reflected in the paired code, whereas it outperforms all baselines on OOD benchmark (MRR of 0.33).
Notably, these improvements are achieved by reformulating the retrieval objective rather than increasing model scale.
We analyze the failure cases on in-distribution set in Section~\ref{sec:our_fail}.

\begin{table}[t]
    \centering
    \small
    \caption{Efficiency Comparison.}
    \label{tab:runtime}
    \resizebox{0.8\linewidth}{!}{
    \begin{tabular}{lccc}
    \toprule
    \textbf{Metric} & \textbf{GraphCodeBERT \cite{graphcodebert}} & \textbf{Qwen2.5-Coder-7B-lt-SupCon-CSN \cite{chen2024decoder}} & \textbf{\ourmethod} \\
    \midrule
    \textbf{Parameters} & 125M & 7B & 125M \\
    \textbf{Inference (per query)} & 0.01s & 0.10s & 0.03s \\
    \textbf{GPU Memory} & 2.1 GB & 28.5 GB & 2.3 GB \\
    \bottomrule
    \end{tabular}
    }
\end{table}


\subsection{Case Studies}

\begin{figure}[t]
    \centering
    \includegraphics[width=0.8\linewidth]{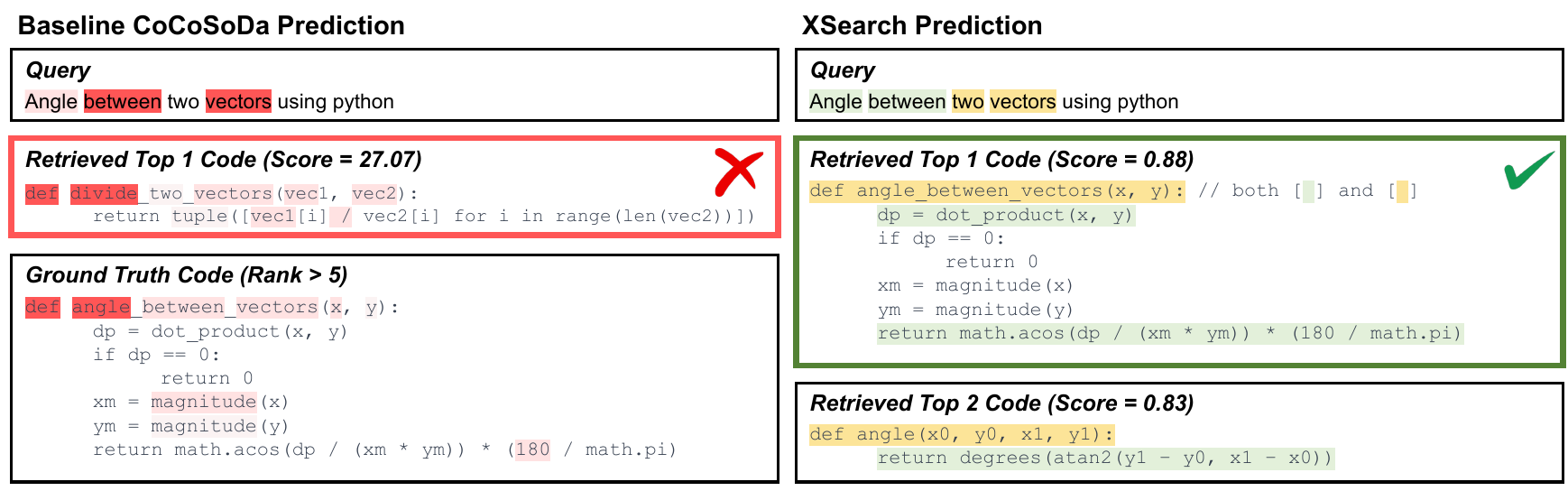}
    \caption{Failure Case of CoCoSoDa \cite{shi2023cocosoda}.}
    \label{fig:cocosoda_fail}
\end{figure}

\subsubsection{\textbf{Why does \ourmethod outperform encoder-based baselines?}}
Figure~\ref{fig:cocosoda_fail} shows a failure case of the strongest encoder-based baseline, CoCoSoDa.
The failure reason is similar to that of CodeBERT in Section~\ref{sec:motivation-eg}.
The query asks for computing the \textit{angle} between two vectors.
However, CoCoSoDa retrieves a code snippet that performs \textit{division} between two vectors, which is lexically similar to the query as both mention ``two vectors'',
but is functionally irrelevant.
This happens because bi-encoder retrieval optimizes similarity between single \texttt{[CLS]} representations, 
without enforcing coverage of all query concepts.
As a result, representations may be dominated by salient or frequent tokens (e.g., ``two vectors''), leading to incorrect matches.
In contrast, \ourmethod ranks the ground-truth implementation as the Top-1 result with a high similarity score. 
The Top-2 result is also semantically correct, except that the two vectors are expressed in coordinate form.
We show more examples on our anonymous website \cite{xsearch}.

\subsubsection{\textbf{Why does \ourmethod outperform decoder-based baselines?}}

\begin{figure}[t]
    \centering
    \includegraphics[width=0.8\linewidth]{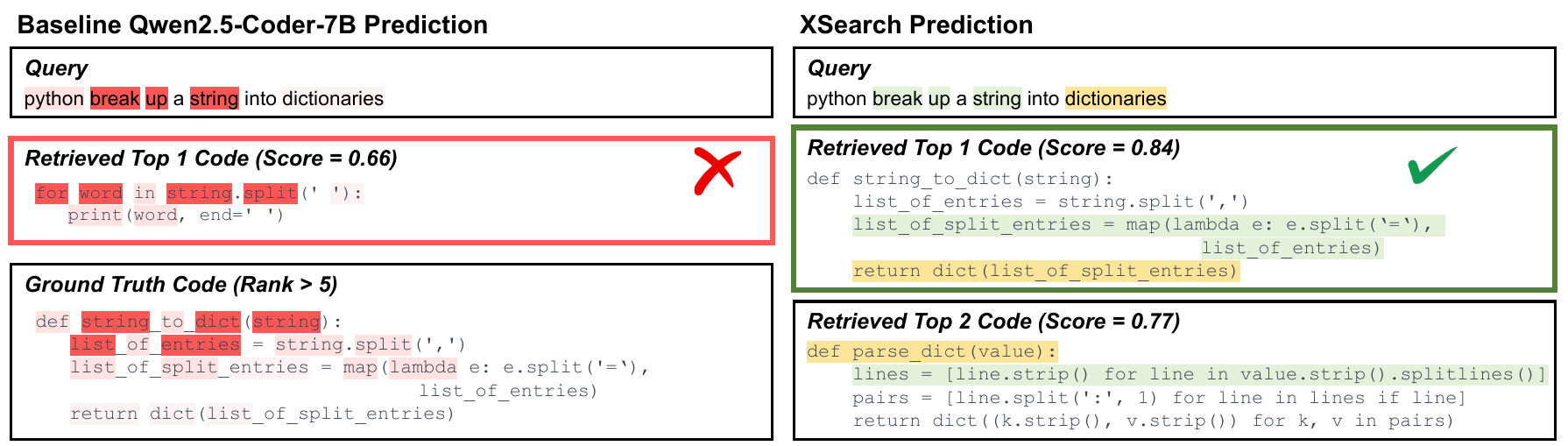}
    \caption{Failure Case of Qwen2.5-Coder-7B-lt-SupCon-CSN  \cite{chen2024decoder}.}
    \label{fig:decoder_fail}
\end{figure}

Despite the strong reasoning capabilities of large decoder-based models, \autoref{fig:decoder_fail} demonstrates that they can still overlook specific functional constraints when processing OOD queries.
In this example, the query requires transforming a string into a \textit{dictionary} object. 
However, the baseline (Qwen2.5-Coder-7B-lt-SupCon-CSN ) retrieves code that merely splits and prints the string, failing to construct the required dictionary structure.
This behavior can be attributed to the representation formulation used in decoder-based retrieval, where a sequence is represented by the average over all token embeddings~\cite{chen2024decoder}.
Under this formulation, the contributions of constraint-bearing tokens can be diluted by other tokens in the sequence.
To verify this effect, we compute the embedding shift induced by masking individual query tokens.
We observe that masking the core constraint “dictionaries” leads to only a negligible change in the averaged embedding.
This indicates that the ``dictionaries'' influence is suppressed by the pooling operation, and the final representation is dominated by more generic action-related tokens such as ``break up''.
In contrast, \ourmethod avoids this dilution effect by computing concept alignments independently, ensuring that every concept directly influences retrieval decisions.

\subsubsection{\textbf{Why does \ourmethod slightly under-perform on in-distribution testing?}}\label{sec:our_fail}

\begin{figure}[t]
    \centering
    \includegraphics[width=\linewidth]{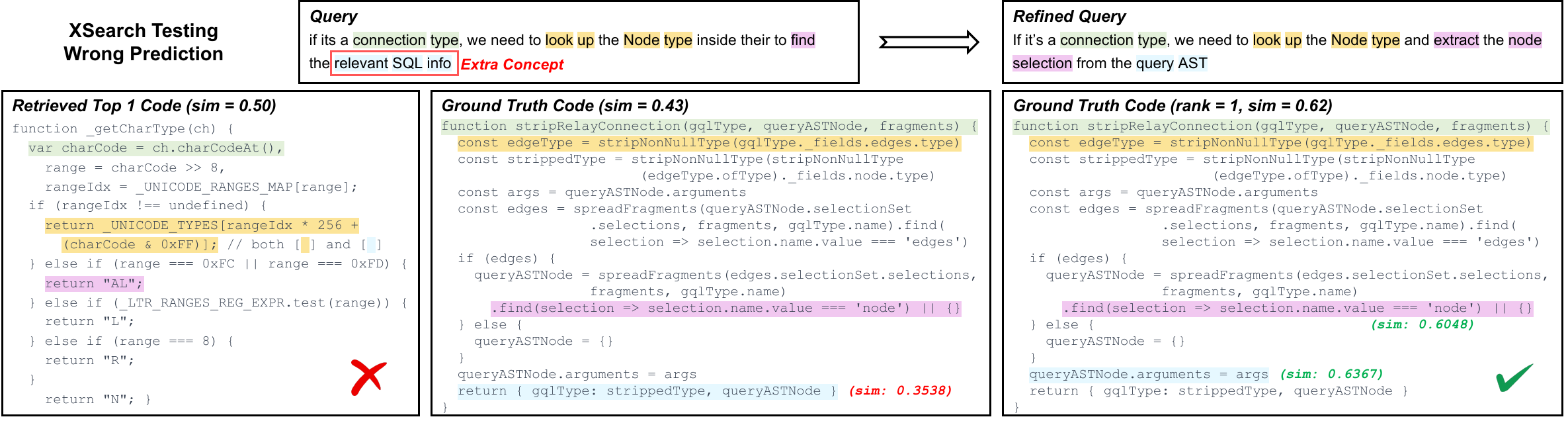}
    \caption{Failure Case of XSearch. 
    \textit{Left:} Retrieved Top-1 code for original query.
    \textit{Middle:} Ground-truth code for original query, where the query includes an extra concept (``relevant SQL info'') not reflected in the code.
    \textit{Right:} Refined query (replacing the extra concept with ``relevant query AST'') and the ground-truth code.}
    \label{fig:xsearch_fail}
\end{figure}
To clarify the ID performance gap, we analyzed 93 failure cases and found that the most prevalent error (32.3\%) stems from redundant query concepts that lack implementation in the code. 
Other main factors include single-concept collapse (22.6\%), imperfect alignment (15.1\%), and missed concepts (11.8\%).
\ourmethod adopts a stricter matching criterion than prior retrievers. 
While most baselines optimize a single global query-code similarity where partial semantic overlap can yield a high score, \ourmethod emphasizes \textbf{functional requirement completeness} by requiring coverage of all concepts in the query. 
As a result, when a query includes extra or overly specific functional-requirement concepts that are not implemented in the corresponding code, 
\ourmethod will intentionally assign a lower similarity score rather than rewarding partial matches.
\autoref{fig:xsearch_fail} illustrates this leading error mode (the 32.3\% category) through a concrete example. 
The original query includes an extra concept, ``relevant SQL info'', which is not reflected in the ground-truth code, 
indicating that the query and code are not synchronized in their information content. 
This missing functional requirement concept reduces the overall query-code score, and the top retrieved result is also low-scored, suggesting limited reliability. 
For developers, returning a high-scored snippet that fails (or cannot be verified) to satisfy an explicitly stated functional requirement can be more misleading than returning no confident result.
When refining the query by replacing the extra concept with a code-grounded description (``extract the node selection from the query AST''), the ground-truth code aligns with all query concepts and achieves a substantially higher similarity. 
More examples are available on our anonymous website \cite{xsearch}.

%% file: tables/table_id_and_ood.tex
\begin{table*}[t]
\centering
\caption{Retrieval Performance Comparison.
Best Scores are Shown in \textbf{Bold}.}
\label{tab:generalization_performance}
\resizebox{\linewidth}{!}{
\begin{tabular}{lcccccccc}
\toprule
\multirow{2}{*}{\textbf{Model}} &
\multicolumn{6}{c}{\textbf{\makecell{In-Distribution \\ (CodeSearchNet)}}} &
\multicolumn{2}{c}{\textbf{\makecell{Out-of-Distribution \\ (CoSQA$^+$)}}} \\
\cmidrule(lr){2-7} \cmidrule(lr){8-9}
& \textbf{\makecell{Ruby \\ (MRR)}} & \textbf{\makecell{JavaScript \\ (MRR)}} & \textbf{\makecell{Go \\ (MRR)}} & \textbf{\makecell{Python \\ (MRR)}} & \textbf{\makecell{Java \\ (MRR)}} & \textbf{\makecell{PHP \\ (MRR)}}
& \textbf{\makebox[1.5cm]{MRR}} & \textbf{MMRR} \\
\midrule
CodeBERT~\cite{feng2020codebert}          & 0.672 & 0.621 & 0.882 & 0.672 & 0.677 & 0.626 & 0.018 & 0.012 \\
UniXCoder~\cite{guo2022unixcoder}         & 0.752 & 0.683 & 0.917 & 0.723 & 0.724 & 0.674 & 0.032 & 0.022 \\
GraphCodeBERT~\cite{guo2020graphcodebert} & 0.703 & 0.644 & 0.897 & 0.692 & 0.691 & 0.649 & 0.020 & 0.010 \\
CoCoSoDa~\cite{shi2023cocosoda}           & \textbf{0.803} & 0.748 & 0.923 & 0.742 & \textbf{0.753} & 0.681 & 0.028 & 0.022 \\
HedgeCode~\cite{chen2024hedgecode}         & 0.785 & 0.748 & 0.883 & 0.718 & 0.731 & 0.686 & 0.026 & 0.020 \\
RAPID~\cite{fan2024rapid}         & 0.745 & \textbf{0.767} & \textbf{0.925} & 0.739 & 0.726 & 0.660 & 0.054 & 0.036 \\
CodeLlama-7B-lt-SupCon-CSN~\cite{chen2024decoder}         & 0.569 & 0.645 & 0.747 & 0.748 & 0.742 & 0.686 & 0.163 & 0.114 \\
Qwen2.5-Coder-7B-lt-SupCon-CSN~\cite{chen2024decoder}         & 0.645 & 0.656 & 0.772 & \textbf{0.793} & 0.754 & \textbf{0.688} & 0.240 & 0.178 \\
\midrule
\textbf{Ours}                             & 0.725 & 0.654 & 0.892 & 0.720 & 0.690 & 0.604 & \textbf{0.332} & \textbf{0.242} \\
\bottomrule
\end{tabular}
}
\end{table*}

%% file: experiments/rq2.tex
\subsection{\textbf{RQ2: Faithfulness of Concept Explanations}}

\begin{figure*}[htbp]
    \centering
    \includegraphics[width=0.8\textwidth]{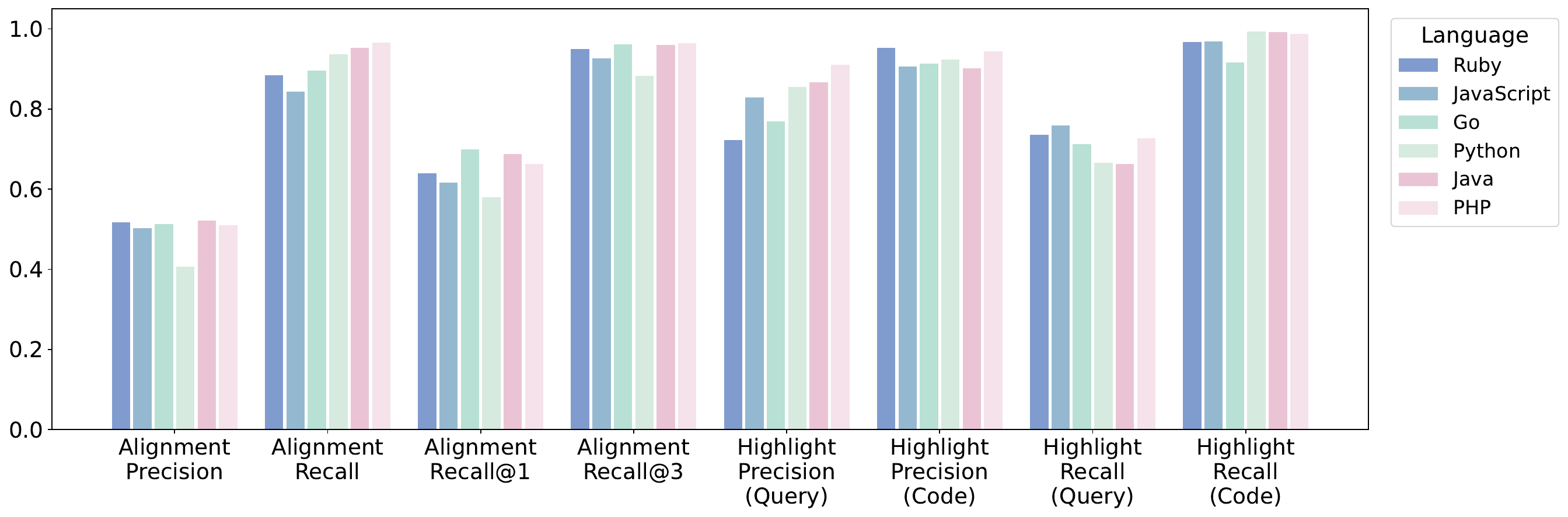}
    \caption{Alignment and Highlight Performance Across Languages.}
    \label{fig:performance-metrics}
\end{figure*}

\subsubsection{\textbf{Setup}}


We define two metrics for assessing the accuracy of two explanation components
produced by XSearch: (i) the predicted concept-bearing tokens and (ii) the predicted concept-to-code alignments.
\begin{itemize}[leftmargin=*]
    \item \textbf{Highlight Accuracy.}
    We use $\delta_{highlight}$ to threshold whether a token is predicted as concept-bearing. 
    We then measure the \textbf{Precision} and \textbf{Recall} of these predicted tokens, using our annotated labels as the ground truth. 
    The final evaluation metric is the average precision and recall across CodeSearchNet testing split.

    \item \textbf{Alignment Accuracy.}
    To evaluate alignment, we compute the cosine similarity between each 
    query concept centroid (Equation~\ref{eq:query-concept-cluster}) and 
    all code line embeddings (Equation~\ref{eq:code-line}). 
    If a code line's similarity exceeds a predefined threshold $\delta_{align}$, it is considered an aligned candidate. 
    We then measure the average alignment \textbf{Precision} and \textbf{Recall} by comparing whether the predicted aligned code lines are within the annotated ground-truth alignments. 
    Additionally, we report \textbf{Recall@k}, 
    which evaluates whether the top-k most similar code lines retrieved for each query concept token include any of the annotated ground-truth alignments. 
\end{itemize}

\subsubsection{\textbf{Results}}

As shown in \autoref{fig:performance-metrics}, \ourmethod achieves consistently high performance across all six programming languages in both alignment and highlight evaluation.

For concept-to-code alignment, the model achieves high top-3 recall (Alignment Recall@3 $> 0.88$), 
meaning that the correct code statements are usually ranked among the top candidates. 
For concept highlighting, both precision and recall are high for query and code tokens, 
indicating that the model reliably identifies tokens that correspond to meaningful concepts, in line with human annotations.

Taken together, these results indicate that the explanations produced by \ourmethod are faithful to the ground truth: 
both highlighted tokens and aligned code statements closely match the ground-truths. 
This level of faithfulness supports the use of concept alignment as a reliable explanation mechanism and provides basis for our user study in Section~\ref{sec:user-study}.


%% file: experiments/rq3.tex
\subsection{\textbf{RQ3: Ablation Study}}

\subsubsection{\textbf{Setup}}
We conduct ablation studies to assess the contributions of alignment loss and highlight loss. 
We consider the following variants:
\begin{itemize}[leftmargin=*]
    \item \textbf{Alignment loss only:} 
    The model is trained with only the alignment loss without highlight loss, i.e. all tokens are assumed to be highlighted.
    \item \textbf{Highlight loss only:} 
    The model is trained with only the highlight loss, refining token importance and highlight distribution.
    \item \textbf{Alignment loss w/o negative sampling + Highlight loss:} 
    The model is trained on both losses, but all negatives participate in alignment loss computation.
    \item \textbf{Alignment loss w/o cross-sample negatives + Highlight loss:} 
    The model is trained on both losses, but all negatives come from the intra-sample code (Section \ref{sec:alignment-loss}).
    \item \textbf{Alignment loss + Highlight loss w/o type embedding:} 
    The model is trained on both losses, but without adding data type embeddings to the code tokens' hidden states.
    \item \textbf{Full setting:} 
    The model is trained with both alignment loss and highlight loss, evaluating their combined effect on overall performance.
\end{itemize}

And we observe the MRR performance of different settings.

\input{tables/table_ablation_study}


\subsubsection{\textbf{Results}}

\autoref{tab:ablation_study} summarizes the ablation results across six programming languages. 
The full setting, which combines alignment loss, highlight loss, cross-sample negative sampling, and type embeddings, consistently achieves the best performance. 
Removing any component leads to a clear and consistent drop in retrieval accuracy.

Using alignment loss alone degrades performance, suggesting that alignment benefits from explicit supervision on which tokens correspond to meaningful concepts. 
In contrast, relying only on highlight loss results in an even larger decline, indicating that identifying important tokens without enforcing their correspondence to code is insufficient for effective retrieval.

We also observe a substantial performance drop when cross-sample negative sampling is removed from the alignment loss. 
This suggests that hard negatives drawn across samples play an important role in preventing mis-alignments. 
Overall, these results show that all the components of \ourmethod work together in a complementary manner.




%% file: tables/table_ablation_study.tex
\newcommand{\cmark}{\ding{51}} 
\newcommand{\xmark}{\ding{55}} 

\begin{table*}[t]
\centering
\caption{Ablation Study of \ourmethod. Best scores are in \textbf{bold}.}
\label{tab:ablation_study}
\resizebox{0.8\linewidth}{!}{
\begin{tabular}{@{}cccccccccc@{}}
\toprule
\multicolumn{4}{c}{\textbf{Components}} & \multicolumn{6}{c}{\textbf{Languages (MRR)}} \\
\cmidrule(lr){1-4} \cmidrule(lr){5-10}
\makecell{\textbf{Alignment Loss}\\\textbf{(In-sample Neg.)}} &
\makecell{\textbf{Alignment Loss}\\\textbf{(Cross-sample Neg.)}} &
\makecell{\textbf{Highlight}\\\textbf{Loss}} &
\makecell{\textbf{Type}\\\textbf{Emb.}} &
\textbf{Ruby} & \textbf{JS} & \textbf{Go} & \textbf{Python} & \textbf{Java} & \textbf{PHP} \\
\midrule
\cmark & \cmark & \xmark & \xmark & 0.647 & 0.586 & 0.860 & 0.687 & 0.672 & 0.511 \\
\xmark & \xmark & \cmark & \xmark & 0.002 & 0.008 & 0.003 & 0.004 & 0.003 & 0.001 \\
\xmark & \cmark & \cmark & \xmark & 0.636 & 0.578 & 0.847 & 0.689 & 0.646 & 0.513 \\
\cmark & \xmark & \cmark & \xmark & 0.017 & 0.034 & 0.024 & 0.012 & 0.090 & 0.022 \\
\cmark & \cmark & \xmark & \cmark & 0.668 & 0.577 & 0.857 & 0.701 & 0.667 & 0.551 \\
\cmark & \cmark & \cmark & \cmark &
\textbf{0.725} & \textbf{0.654} & \textbf{0.892} &
\textbf{0.720} & \textbf{0.690} & \textbf{0.604} \\
\bottomrule
\end{tabular}
}
\end{table*}

%% file: experiments/rq4.tex
\subsection{\textbf{RQ4: User Study on Explanation}}\label{sec:user-study}

\begin{figure*}[htbp]
    \centering
    \begin{subfigure}{.48\linewidth}
    \centering
        \includegraphics[scale=0.3]{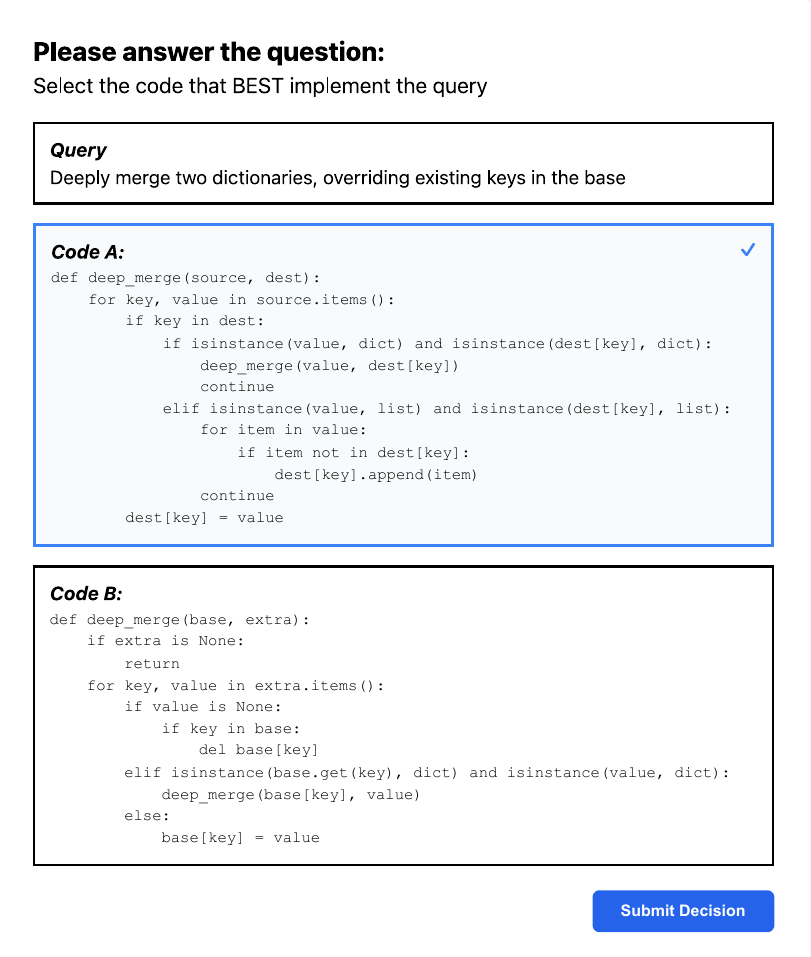}
        \caption{Control Group Interface}\label{fig:ui_cg}
    \end{subfigure}
        \begin{subfigure}{.48\linewidth}
        \centering
        \includegraphics[scale=0.3]{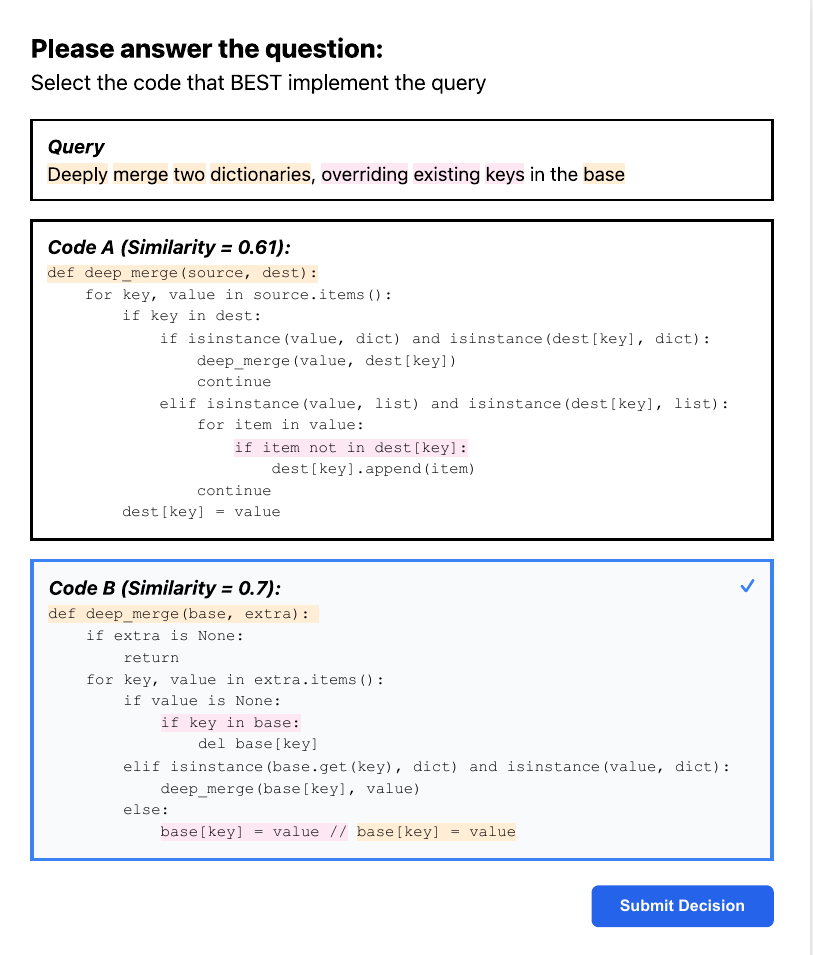}
        \caption{Experimental Group Interface}\label{fig:ui_eg}
    \end{subfigure}
    \vspace{-10pt}
    \caption{
    The query asks for an in-place merge of two dictionaries and overwrites existing keys in the base dictionary.
    Code A didn't overwrite existing keys, so the correct answer is Code B.
    Without explanations, users might omit such details and select the wrong answer.
    }
    \label{fig:user_study_example}
\end{figure*}

We conduct a controlled user study to test whether \ourmethod's concept-alignment explanations help developers make (i) \textbf{faster} and (ii) more \textbf{accurate} accept or reject decisions when choosing among plausible code candidates.


\subsubsection{\textbf{Participants}}
We recruited 20 participants from four top universities with different backgrounds, including undergraduate students, master's students, PhD candidates, and research staff. 
Before the study, each participant was asked to complete a short questionnaire covering their (i) academic level, (ii) years of programming, and (iii) self-reported Python familiarity.
We then formed 10 matched pairs with similar profiles and assigned one participant to the \textbf{experimental group (EG)} and the other to the \textbf{control group (CG)}.
As a result, two groups are of equal size ($n=10$), with the same demographic distribution.



\subsubsection{\textbf{Task Design}}
We constructed 20 evaluation tasks from CodeSearchNet in \textbf{Python}. 
Each task paired a natural-language query with five candidates, only one of which matched the query intent. The 4 distractors were selected from the top-4 retrieved results of GraphCodeBERT that were \textit{not} the ground truth (i.e., plausible but functionally incorrect). The single correct answer for each task was manually verified by two authors independently.
Tasks spanned five functional categories (Table~\ref{tab:task_distribution}) and three difficulty levels based on Line-of-Code (LOC) and branch complexity: Easy ($<8$ LOC), Medium ($8$–$16$ LOC), and Difficult ($>16$ LOC).

\begin{table}[ht]
\centering
\caption{Task Distribution by Functional Category and Examples}
\label{tab:task_distribution}
\scriptsize
\begin{tabular}{lcp{18em}}
\toprule
\textbf{Category} & \textbf{Count} & \textbf{Examples} \\ 
\midrule
System and Configuration Management & 5 & \textit{Removes an environment variable} \\
\hline
Data Conversion & 5 & \textit{Convert any value to a boolean} \\
\hline
Data Parsing & 4 & \textit{Extract the parts of a date given a timestamp} \\
\hline
Data I/O & 4 & \textit{Writes the experimental setup to a JSON file} \\
\hline
Data Validation & 2 & \textit{Check allowed file extensions}
\\ 
\midrule
\textbf{\textit{Total}} & 20 & -- \\
\bottomrule
\end{tabular}
\end{table}


\subsubsection{\textbf{UI Interface}}
The CG interface (Figure~\ref{fig:ui_cg}) mimics standard search, displaying only candidate code. 
The EG interface (Figure~\ref{fig:ui_eg}) additionally presents \ourmethod's explanations: (i) highlighted query concepts and (ii) their corresponding aligned code spans in each candidate. Candidate ordering and ground-truth answers remained identical across groups.

\subsubsection{\textbf{Study Setup}}
Before starting the main session of the user study, all participants complete a warm-up session consisting of 5 practice tasks. 
This session allows participants to familiarize themselves with the task format and interface without contributing to the evaluation results. 
Both groups receive the same warm-up tasks to ensure equal preparation.
In the main session, we evaluate the participants' performance along two perspectives: 
\begin{itemize}[leftmargin=*]
    \item \textbf{Accuracy}, i.e., whether the participant selects the correct code snippet that matches the query's intended functionality;
    \item \textbf{Completion Time}, i.e., the time taken to make a selection for the main session.
\end{itemize}


To compare the performance between the CG and EG, we use the Mann-Whitney U test \cite{mann1947test}. This test is appropriate for independent samples when the assumption of normality may not hold, and it evaluates whether there is a significant difference in the distributions of the two groups.

\subsubsection{\textbf{Results}}
The results are summarized in \autoref{tab:user_expertise}, with the following observation:

\begin{itemize}[leftmargin=*]
    \item \textbf{Accuracy:} The experimental group (EG), achieves an average accuracy of \textbf{80.00\%}, compared to \textbf{69.50\%} in the control group (CG), showing an improvement of \textbf{10.5\%}. 
    The Mann-Whitney U test reveals that this improvement is statistically significant ($p = 0.0297 < 0.05$), indicating that \ourmethod substantially improves accuracy.

    \item \textbf{Efficiency:} The EG group completes the tasks faster, with an average task time of \textbf{1775.2} seconds compared to \textbf{2863.4} seconds in the CG group, resulting in a \textbf{38\%} reduction in time. 
    The Mann-Whitney U test also shows this difference is statistically significant ($p = 0.0188 < 0.05$), indicating a clear efficiency gain from the enhanced interface.

\end{itemize}

\input{tables/table_user_study}

\noindent\textbf{How highlighting helps users?} 
Our results confirm that concept highlighting significantly reduces cognitive load across all users' programming experience levels. 
Notably, we observed a \textbf{leveling up} effect among \textbf{Intermediate} users.
With \ourmethod's explanations, they achieved higher accuracy than Expert users in the control group (85\% vs. 66\%) while requiring less than half the time (1300s vs. 2800s). This suggests the tool effectively bridges the expertise gap by directing attention to critical logic instead of requiring line-by-line reading.

 
\noindent\textbf{Why does EG achieve higher accuracy?} 
Post-study interviews with CG (Control Group) participants revealed that incorrect choices were primarily due to \textbf{keyword match} bias. 
Participants often chose an incorrect candidate simply because it contained some keywords similar to the query, despite missing the core logic. 
As shown in the case in Figure~\ref{fig:user_study_example}, the correct and incorrect snippets share significant keyword overlaps. 
Without the concept-to-line alignments provided by \ourmethod, CG participants failed to identify where the logic differs from the query's intent 
(e.g., Only Code B implements "overriding" existing keys).

%% file: tables/table_user_study.tex


\begin{table}[h]
\centering
\caption{Performance Comparison: Novice vs. Expert Groups}
\label{tab:user_expertise}
\scriptsize
\begin{tabular}{lcccccc}
\toprule
\textbf{User Experience Level} & \multicolumn{2}{c}{\textbf{Accuracy (\%)}} & \multicolumn{2}{c}{\textbf{Time (s)}} & \multicolumn{2}{c}{\textbf{Improvement}} \\
\cmidrule(lr){2-3} \cmidrule(lr){4-5} \cmidrule(lr){6-7}
& \textbf{CG} & \textbf{EG} & \textbf{CG} & \textbf{EG} & \textbf{$\Delta$ Acc} & \textbf{Time Saved} \\
\midrule
Low (Novice)         & 75.00 & 76.25 & 2811.7 & 2162.3 & +1.25  & 23.1\% \\
Medium (Intermediate)& 45.00 & 85.00 & 3336.0 & 1300.0 & \textbf{+40.00} & \textbf{61.0\%} \\
High (Expert)        & 66.67 & 82.00 & 2809.3 & 1560.6 & +15.33 & 44.5\% \\
\midrule
\textbf{Average}     & \textbf{69.50} & \textbf{80.00} & \textbf{2863.4} & \textbf{1775.2} & \textbf{+10.5\%} & \textbf{38.0\%} \\
\bottomrule
\end{tabular}
\end{table}

%% file: threat.tex
\section{Discussion}
\label{sec:discussion}

\noindent\textbf{Faithfulness and Usefulness of Explanations.} 
The value of \ourmethod lies in both the faithfulness and the practical usefulness of its explanations. For faithfulness, as demonstrated in Figure~\ref{fig:performance-metrics}, the concept-to-code alignments themselves serve as intrinsic explanations, achieving over 80\% Recall and 88\% Recall@3 across six languages. This quantitatively confirms that the generated rationales accurately reflect the ground truth. For usefulness, our controlled user study (Section~\ref{sec:user-study}) grounds the evaluation in the actual downstream application: developers selecting the best-fit code. The results show that with \ourmethod's explanations, developers are 38\% faster and 10.5\% more accurate, proving that alignment explanations effectively reduce cognitive load and keyword-match bias.

\noindent\textbf{Impact of the Supervision Pipeline (Pseudo-label Ablation).} 
To understand the extent to which \ourmethod's improvements depend on the LLM annotation pipeline, we conducted an ablation study using automatically derived pseudo-labels. We clustered token embeddings from a pretrained GraphCodeBERT to form query concepts, used per-line average embeddings as code units, and aligned them via nearest-centroid matching. The \ourmethod architecture and training objectives remained identical. On the CoSQA+ out-of-distribution benchmark, the pseudo-label variant achieved an MRR of 0.136, significantly outperforming the unsupervised GraphCodeBERT baseline (0.020). However, it fell short of the 0.332 MRR achieved with LLM labels. This indicates that \ourmethod's performance gains stem from two synergistic factors: the deductive alignment architecture intrinsically improves generalization (architecture contributes), and the high-quality LLM supervision further unlocks the model's potential (supervision contributes).

\section{Threats to Validity}

\noindent\textbf{Internal threats.}
Concept alignment labels are generated using GPT-4o~\cite{gpt4o} and may contain semantic noise.
We mitigate this risk through formatting constraints and manual verification, though residual noise may remain.
Our goal is not perfect annotation, but to have richer supervision through label augmentation.
Model performance may also depend on hyperparameter choices.
However, our sensitivity analysis (Figure~\ref{fig:hyperparam_sweep}) shows stable performance within reasonable ranges.

\noindent\textbf{External threats.}
Our evaluation is based on CodeSearchNet and CoSQA+, which may not cover all real-world code search scenarios, such as large repository-level retrieval.
Nevertheless, these benchmarks span six languages and are widely used in prior works, enabling fair comparison.
Our user study involves a limited number of participants under controlled settings, further real-world studies are needed to assess long-term usability.

%% file: conclusion.tex
\section{Conclusion}

In this work, we present \ourmethod, an intrinsically explainable code retrieval model that addresses two key limitations of existing approaches: limited explainability and poor generalization under distribution shift. 
Unlike traditional methods that rely on global embedding search, \ourmethod explicitly formulates retrieval as a concept-alignment task. 
This reframes retrieval from an inductive to a deductive perspective: the model is designed to (i) identify the functional requirements in a query and (ii) verify how many of those requirements are satisfied by a candidate code snippet. 
Extensive experiments show that \ourmethod achieves competitive performance on in-distribution benchmarks and outperforms state-of-the-art models by up to 15× on out-of-distribution datasets. 
In addition, our user study demonstrates that concept-alignment explanations improve decision speed by 38\% and accuracy by 10.5\% when users accept or reject code recommendations. 
Overall, \ourmethod bridges the gap between performance and interpretability in code search, offering a scalable, user-centric solution for semantic retrieval.

\section{Acknowledgment}
This research is supported in part by the National Natural Science Foundation of China (62572300), the Minister of Education, Singapore (MOE-T2EP20124-0017, MOET32020-0004), the National Research Foundation, Singapore and the Cyber Security Agency under its National Cybersecurity R\&D Programme (NCRP25-P04-TAICeN), DSO National Laboratories under the AI Singapore Programme (AISG Award No: AISG2-GC-2023-008-1B), and Cyber Security Agency of Singapore under its National Cybersecurity R\&D Programme and CyberSG R\&D Cyber Research Programme Office, and partially by HUAWEI’s Al Hundred Schools Program using the Ascend AI technology stack. Any opinions, findings and conclusions or recommendations expressed in this material are those of the author(s) and do not reflect the views of National Research Foundation, Singapore, Cyber Security Agency of Singapore as well as CyberSG R\&D Programme Office, Singapore.